\documentclass[%
preprint,
amsmath,amssymb,
aps,
pre,
]{revtex4-1}

\usepackage{hyperref} 

\usepackage{graphicx} 
\usepackage{dcolumn}  
\usepackage{bm}       

\usepackage{graphics}
\usepackage{color}

\providecommand{\abs}[1]{\vert #1\vert}

\newcommand{\const}{\mathop{}\!\mathrm{const}}

\begin{document}

\title{Numerical integration of KPZ equation with restrictions}
\author{M. F. Torres}
\email{mtorres@ifimar-conicet.gob.ar}
\author{R. C. Buceta}
\email{rbuceta@mdp.edu.ar}
\affiliation{Instituto de Investigaciones F\'{\i}sicas de Mar del Plata (UNMdP and CONICET)}
\affiliation{Departamento de F\'{\i}sica, FCEyN, Universidad Nacional de Mar del Plata \\ Funes 3350, B7602AYL Mar del Plata, Argentina}
\date{\today}

\begin{abstract}
In this paper, we introduce a novel integration method of Kardar-Parisi-Zhang (KPZ) equation. It has always been known that if during the discrete integration of the KPZ equation the nearest-neighbor height-difference exceeds a critical value, an instability appears and the integration diverges. One way to avoid these instabilities is to replace the KPZ nonlinear-term by a function of the same term that depends on a single adjustable parameter which is able to control pillars or grooves growing on the interface. Here, we propose a different integration method which consists of directly limiting the value taken by the KPZ nonlinearity, thereby imposing a restriction rule that is applied in each integration time-step, as if it were the growth rule of a restricted discrete model, {\sl e.g.} restricted-solid-on-solid (RSOS). Taking the discrete KPZ equation with restrictions to its dimensionless version, the integration depends on three parameters: the coupling constant $g$, the inverse of the time-step $k$, and the restriction constant $\varepsilon$ which is chosen to eliminate divergences while keeping all the properties of the continuous KPZ equation. We study in detail the conditions in the parameters' space that avoids divergences in the 1-dimensional integration and reproduce the scaling properties of the continuous KPZ with a particular parameter set. We apply the tested methodology to the $d$-dimensional case ($d = 3,4$) with the purpose of obtaining the growth exponent $\beta$, by establishing the conditions of the coupling constant $g$ under which we recover known values reached by other authors, in particular for the RSOS model. This method allows us to infer that $d = 4$ is not the critical dimension of the KPZ universality class, where the strong-coupling phase dissapears.
\end{abstract}

\pacs{Valid PACS appear here}

\maketitle

\section{Introduction}

Throughout the last three decades, the research of numerous topics on models and equations in order to understand the phenomenon of surface growth has been very intense. Among the systems that grow out of equilibrium, perhaps the most studied ones are those included in the Kardar-Parisi-Zhang (KPZ) universality class characterized by the homonymous stochastic equation \cite{Kardar1986}
\begin{equation}
\frac{\partial h}{\partial t}=\nu\,\nabla^2 h  + \frac{\lambda}{2}\,(\nabla h)^2 + \eta ({\bf x},t)\;,
\label{eq:KPZ}
\end{equation}
where $h=h({\bf x},t)$ is the surface height of a growing medium on a $d$-dimensional substratum, at position ${\bf x}$ and time $t$.  The Laplacian term and the non-linear term represent the elasticity and the lateral growth of the interface, respectively. The noise $\eta ({\bf x},t)$ is Gaussian with zero mean and covariance \mbox{$\langle \eta ({\bf x},t)\eta ({\bf x}',t')\rangle =2 D\,\delta^{d}({\bf x}-{\bf x}')\,\delta(t-t')$}, where $D$ is the noise intensity. The right-hand side of the KPZ equation~(\ref{eq:KPZ}) may include an additive term representing a constant force $f$ due to incoming or outcoming particle-flow that is absorbed or desorbed on the surface, respectively. For $\lambda=0$ the Eq.~(\ref{eq:KPZ}) becomes the Edward-Wilkinson (EW) equation \cite{Edwards1982}. A large amount of real growing interfaces has been successfully described by the KPZ equation in one and two dimensions \cite{Barabasi1995, Krug1997, Krim1995, Miettinen2005}.  Those growing lattice models that have the same interface properties as the KPZ equation, {\sl e.g.} ballistic deposition \cite{Vold1959}, restricted-solid-on-solid (RSOS) \cite{Kim1989}, etching algorithm \cite{Mello2001} or Eden model \cite{Eden1961}, are also used as a substitute for addressing open questions. The existence or not of an upper critical dimension $d_\mathrm{u}$ above of which fluctuations are negligible, independently of the KPZ nonlinearity strength, is currently the most salient unresolved issue \cite{Tang1990, Frey1994, Healy1990, Lassing1997, *Lassing1998, Bouchaud1993, *Bouchaud1993e, Bhattacharjee1998, Colaiori2001, Katzav2002, Fogedby2006, Moore1995, Castellano1998a, Castellano1998b, Castellano1999, AlaNissila1993, Marinari2000, Marinari2002, Pagnani2013, Pagnani2015, Kim2013, Kim2014, Alves2014, Rodrigues2015, Alves2016}.
To consider that the KPZ equation is a proper hydrodynamic description of the interface growth of some real system or growth model, they both need to have the same scaling properties and exponents. The KPZ equation has Galilean invariance, regardless of the dimension $d$, if the relation $\zeta +z = 2$ is verified \cite{Kardar1986, Barabasi1995, Wio2011}, where $\zeta$ is the (global) roughness exponent and $z$ is the dynamical exponent. For $d = 1$, since the system behavior verifies the fluctuation-dissipation theorem \cite{Kardar1986}, we know that $\zeta = 1/2$. This theorem is not valid for $d>1$ for which the exponents must be calculated in a different way, although an analytical method that allows to know the exponents or scaling properties has not yet been established. Several numerical methods have been developed with similar results in some cases and dissimilar in others. The methods of perturbative renormalization  \cite{Kardar1986, Barabasi1995,Tang1990, Frey1994}, with coupling constant $g=\lambda^2D/\nu^3$, predict that for every dimension $d>2$ there is a critical value $g_{\mathrm{c}}\neq0$ that separates two regions: for \mbox{$g<g_{\mathrm{c}}$} a weak-coupling phase with \mbox{$\zeta=0$} and for \mbox{$g>g_{\mathrm{c}}$} a strong-coupling phase with \mbox{$\zeta\neq 0$}. However, these methods have been unable to obtain exponents in the strong-coupling phase. On the one hand, studies of the mapping of the KPZ equation in the directed polymers with quenched noise \cite{Healy1990, Lassing1997} and self-consistent methods \cite{Bouchaud1993, Moore1995, Bhattacharjee1998, Colaiori2001, Katzav2002, Fogedby2006} predict a critical dimension $d_\mathrm{u}=4$, where the strong-coupling phase disappears. On the other hand, using a non-perturbative renormalization method in the real space, exponents of the strong-coupling phase up to dimension $d=9$ were obtained, without the prediction of a critical dimension  \cite{Castellano1998a, *Castellano1998b, *Castellano1999}. In addition, these exponents are close to those obtained from the simulation of growth models, which share the properties of KPZ equation exactly at $d=1$. Specifically, the RSOS model has been simulated up to dimension $11$, without showing signs of a critical dimension \cite{Kim1989, Mello2001, Pagnani2013, AlaNissila1993, Pagnani2015, Marinari2000, Marinari2002, Kim2013, Kim2014, Alves2014}. The discrete integration of the KPZ equation is another of the methods used to obtain its scaling properties. The KPZ equation is usually numerically integrated following a discretization scheme  like this \cite{Moser1991}:
\begin{eqnarray}
h_j(t + \Delta t) &=& h_j(t) + \sum_{i=1}^{i=d}\Bigl(\nu\,L_j^{[i]} + \frac{\lambda}{2}\,N_j^{[i]}\Bigr)\,\Delta t \nonumber\\
&&+\,\sigma \sqrt{12\,\Delta t}\; R_j(t)\;,\label{eq:KPZ-d1}  
\end{eqnarray}
where  $\Delta t$ is the integration time-step and $h_j$ is the height of interface on the $j$-th lattice point. The noise amplitude is $\sigma \sqrt{12\,\Delta t}$, where $\sigma=\sqrt{2D/(\Delta  x)^d}$, and $R(t)$ is a uniform random variable between $\left[-0.5,0.5 \right]$. The linear term \mbox{$L_j^{[i]}=\bigl(h_{j+1}^{[i]}-2\,h_j+h_{j-1}^{[i]}\bigr)/(\Delta x)^2$} is the discrete $\partial^2 h/\partial x_i^2$ of the Laplacian, where $\Delta x$ is the integration mesh-step and  $h_{j\pm 1}^{[i]}$ are the heights of nearest-neighbours (NN) of the $j$-th lattice point in the $i$-th direction. The nonlinear term of the KPZ can be discretized by adopting different schemes that are equivalent \cite{Buceta2005}, although the simplest one is the Euler scheme, in which the nonlinear term of Eq.~(\ref{eq:KPZ-d1}) is \mbox{$N_j^{[i]}=\bigl[\bigl(h_{j+1}^{[i]}-h_{j-1}^{[i]}\bigr)/(2\Delta x)\bigr]^2$}. Replacing $h_j=h_0 H_j$, \mbox{$x=x_0\,r$} and \mbox{$t=t_0\tau$} 
in Eq.~(\ref{eq:KPZ-d1}) we obtain the dimensionless equation
\begin{eqnarray}
H_j(\tau + \Delta\tau) &=& H_j(\tau) +  \sum_{i=1}^{i=d}\Bigl(\mathcal{L}_j^{[i]} + \frac{1}{2}\,\mathcal{N}_j^{[i]}\Bigr)\,\Delta\tau \nonumber\\
&&+\, \sqrt{12\,\Delta \tau}\;R_j(\tau)  
\label{KPZ-d2}
\end{eqnarray}
where \mbox{$h_0=\nu /\lambda$\,}, \mbox{$x_0=\sqrt{\nu^{3}/(\sigma\lambda)^2}$\,}, and \mbox{$t_0=[\nu/(\sigma\lambda)]^2$\,}. In the last equation, the linear and nonlinear terms are
\begin{eqnarray}
&&\mathcal{L}_j^{[i]}=\frac{H_{j+1}^{[i]}(\tau)-2\,H_j(\tau)+H_{j-1}^{[i]}(\tau)}{(\Delta r)^2}\;,\\
&&\mathcal{N}_j^{[i]}=\biggl[\frac{H_{j+1}^{[i]}(\tau)-H_{j-1}^{[i]}(\tau)}{2\,\Delta r}\biggr]^{\!2}\;,\label{KPZ-nonlinear}
\end{eqnarray}
respectively.

The KPZ discrete integration presents instabilities that make its simulation diverge quickly \cite{Tu1992,Kim1994}. A deep analysis made by Dasgupta {\sl et al.} \cite{Dasgupta1996, *Dasgupta1997} shows that these instabilities are caused by the uncontrollable growth of pillar or grooves that are intrinsic to discrete versions of equations with nonlinear terms $(\nabla h)^2$, with or without noise. The instabilities cannot be avoided by making the system bigger or using a generalized discretization of the nonlinear term. Even reducing the integration time step only makes the appearance of instabilities less probable.  The instability in the numerical integration of the one-dimensional KPZ equation with $\lambda > 0$ ($\lambda <0$) is associated with grooves (pillars). Starting from a flat interface with a pillar- or groove-perturbation of height $h_\mathrm{p}$ at some point in the mesh, by mean of the numerical integration Dasgupta {\sl et al.} showed that a critical height $h_\mathrm{c}\propto -\lambda^{-1}$ exists. Below the critical value ({\sl i.e.} $|h_{\mathrm{p}}|<|h_{\mathrm{c}}|$) the perturbation is reabsorbed into the interface and while above the critical value the interface diverges quickly. It can observed that if during the discrete integration of the KPZ equation the NN height difference surpasses a critical value, the instability appears and the integration diverges. 
The exact solution to the one-dimensional KPZ equation found by Sasamoto and Spohn \cite{Sasamoto2010} shows that there are no instabilities in the continuous version. Additionally, the  noiseless continuous KPZ equation can be mapped into a diffusion equation by a Cole-Hopf transformation and be solved exactly without any instabilities. Conversely, the application of such a transformation to the noiseless discrete KPZ equation is not reduced to a discrete diffusion equation, which suggests a possible explanation for the generation of instabilities along the numerical integration. 
To successfully integrate the KPZ, Dasgupta et al. \cite{Dasgupta1996, Dasgupta1997} propose to replace the non-linear term $(\nabla h)^2$ with a function $\Phi((\nabla h)^2)$ in the KPZ equation, defined by $\Phi(y)=(1-e^{-cy})/c$\,, where $c$ is an adjustable constant. This proposal does not require a special discretization scheme, since it is possible to maintain the Euler scheme without loss of generality. This method avoids, within a certain range of the parameter $c$, the big local height-differences that lead to excessive growth, which are the origin of instabilities. Introducing the nonlinear function $f$ into the KPZ equation, which is equivalent to introducing an infinite nonlinear series must leave the scaling properties of universal quantities invariant, in addition to eliminating divergences. Some properties of the KPZ equation, such as scaling exponents' can be calculated with this procedure \cite{Dasgupta1997} with great precision and coinciding with the theoretical values from renormalization group for $d=1$ and growth models for $d=2$ \cite{Miranda2008}. It is important to note that if $c\gg 1$, the nonlinear effects become very weak and, in the extreme case, there is a long transient with scaling properties of the EW universality class. On the contrary, if $c\ll 1$, the method fails to avoid instabilities.

Inspired by the study of the instabilities made by Dasgupta et al. \cite{Dasgupta1997}, in this work we propose a different integration model. By our method, we directly limit the value of the non-linear term, restricting the value of height-difference between NN columns, a difference which is responsible for making integration divergent. In this paper, we are going to characterize this method and use it to regain and obtain important properties of the KPZ equation. In Section~\ref{sec:2} we introduce and characterize the restricted integration method. In Subsection~\ref{subsec:3a} we show the integration properties of the KPZ equation in \mbox{$d = 1$} and obtain the associated exponents. In Subsection~\ref{subsec:3b} we show the results for \mbox{$d = 3$}, which was never obtained with an integration scheme that avoids divergences, and for \mbox{$d = 4$}, which is of great importance since the KPZ numerical integration was never performed until this dimension, and because it is the dimension that several authors predict to be  critic, {\sl i.e.} that does not present a roughness region.

\section{Model and definitions\label{sec:2}}
\begin{figure}[h!]
\centering
\includegraphics[scale=0.3]{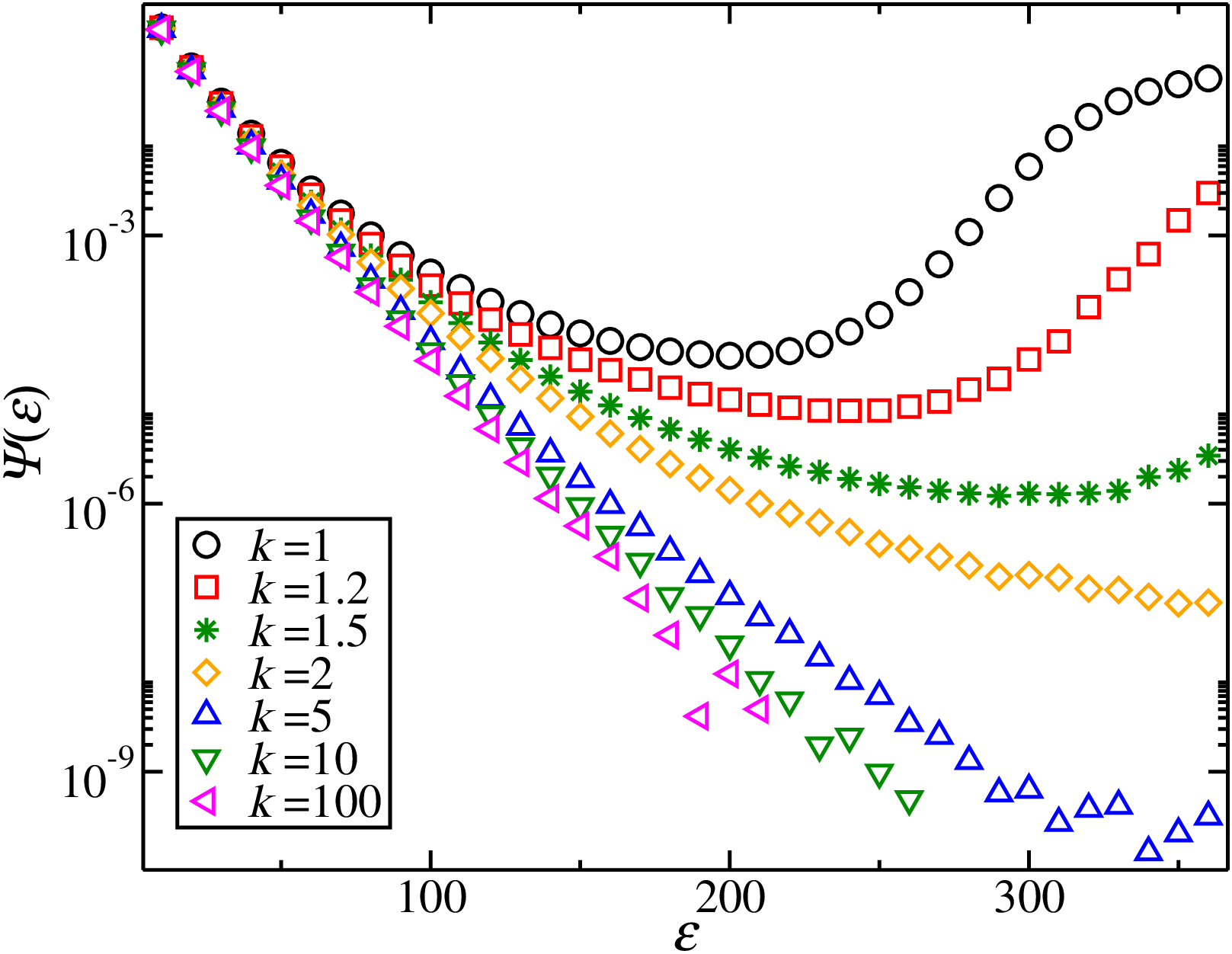}
\caption{(Color online) Plot of the probability $\Psi$ as a function of the restriction constant $\varepsilon$ for several values of $k=(\Delta\tau)^{-1}$, where $\Delta\tau$ is the integration time-step. We take $L=512$ and $g=12.56$. We use the integration mesh-step $\Delta r=\sqrt{2g}$ corresponding to taking $\Delta x=1$.\label{Fig:1}}
\end{figure}

To integrate the KPZ equation we are going to use a similar scheme to the one presented in Eq.~(\ref{KPZ-d2}), but with an upper limit to nonlinearity $\mathcal{N}_j^{[i]}$ given by Eq.~(\ref{KPZ-nonlinear}). Our proposal is to replace this term as follows:
\begin{equation}
\mathcal{N}_j^{[i]}\longrightarrow\frac{1}{(\Delta r)^2} \Bigl[\bigl(\mathcal{N}_j^{[i]}(\Delta r)^2-\varepsilon\bigr)\,\Theta\bigl(\varepsilon-\mathcal{N}_j^{[i]}(\Delta r)^2\bigr)+\varepsilon\Bigr]\,,  
\label{KPZ-d3}
\end{equation}
where $\Theta(x)$ is the Heaviside step function. The restricted growth rule \mbox{$\abs{H_{j+1}^{[i]}(\tau)\!-\!H_{j-1}^{[i]}(\tau)}\!\le\! 2\sqrt{\varepsilon}$} applies for each integration time-step, as if it were the growth rule of a discrete model with restrictions ({\sl e.g.} RSOS). The restriction constant $\varepsilon$ must be chosen so as to eliminate the divergences that arise in the usual integration of the discrete KPZ equation and, at the same time, to maintain all the basic properties of the continuous KPZ equation. From Eq.~(\ref{KPZ-d2}) we can easily see that $\varepsilon$ depends on the integration steps $\Delta r$ and $\Delta\tau $. By means of the coupling constant $g=\lambda^2D/\nu^3=\lambda^2\sigma^2/2\nu^3$, introduced from the KPZ studies by renormalization group theory, it is easy to see that integration depend in turn on the coefficients and the integration steps through equations \mbox{$\Delta r=\sqrt{\frac{2g}{\Delta x^{(d-2)}}}$} and \mbox{$\Delta\tau=\frac{2\nu g}{\Delta x^d}\,\Delta t$\,}. Notice that the integration steps $\Delta r$ and $\Delta\tau$ are invariant under the transformations $\Delta x\to 1$, $\Delta t\to \Delta t/(\Delta x)^2$ and $g\to g/(\Delta x)^{d-2}$. This property allows us to take $\Delta x = 1$ without loss of generality. In addition, by choosing $\Delta t=1/(2k\nu g)$ (where $k$ is a positive real constant) we make the integration simpler by ensuring the same statistic for each $g$ value, \mbox{$\Delta \tau=1/k$}. In short, the integration depends on only three parameters: the coupling constant $g$, the inverse of time step $k$ and the restriction value $\varepsilon $.

In order to obtain the range of $\varepsilon$ values for which the properties of the KPZ equation remain invariant, we will study the probability that the restriction to growth occurs, {\sl i.e.} \mbox{$\Psi(\varepsilon)=P(X>\varepsilon)= \langle\Theta(X-\varepsilon)\rangle$}, where \mbox{$X=\{\frac{1}{2}\,[H_{j+1}^{[i]}(\tau)-H_{j-1}^{[i]}(\tau)]\}^2$}. 

One of the main observables that can be measured in a growing interface, in order to characterize its evolution, is its width or roughness defined by \mbox{$w(\tau)=[\{{\langle h({\bf r},\tau) \rangle}^2-\langle{h({\bf}r,\tau)}^2\rangle\}]^{1/2}$}, where $\langle\cdots\rangle$ is the average over the interface of size $L$ and $\{\cdots\}$ over the different realizations. It is usually found that the system has a power law behavior with $w\propto \tau^{\beta}$ for $\tau\ll\tau_\mathsf{x}$, where \mbox{$\beta=\zeta/z$} is the growth exponent, and  that the system saturates with $w=w_\mathrm{sat}\propto L^{\zeta}$ for $\tau\gg\tau_\mathsf{x}$. Also, it has been found that the crossover time $\tau_\mathsf{x}\propto L^z$. In the case of the KPZ equation, because of its Galilean invariance, it is known that $\zeta=1/2$ and $z=3/2$ for $d=1$ and $z +\beta =2$ for any dimension $d$. 

Another observable measured in a growing interface is the height-difference correlation of $m^\mathrm{th}$-order defined by \mbox{$G_m(\ell,\tau)=\{\langle\abs{H_{j+\ell}^{[i]}(\tau)-H_{j}^{[i]}(\tau)}^m\rangle\}$}, where $\ell$ is the distance between two columns. For growing systems it is expected that this correlation will present a power law behavior \mbox{$G_m\propto \ell^{\,m\zeta_m}$ for $\ell\ll\xi(\tau)$} or take a constant value \mbox{$G_m^\mathrm{\,sat}$ for $\ell\gg\xi(\tau)$}, where $\xi(\tau)$ is the correlation length. Also, $\xi\propto\tau^{1/z}$ for $\tau<\tau_\mathsf{x}$ and equal to
the maximum neighbour distance for $\tau>\tau_\mathsf{x}$ (e.g. $\xi=L/2$
for systems with periodic boundary conditions). When $\zeta_m$ depends on $m$ the correlation shows multiscaling and the system is multi-affine. Otherwise, the correlation  shows single scaling and the system is self-affine. The particular case $m=2$ allows to relate the correlation with the roughness. It is observed that $G_2\propto\tau^{2(\zeta-\zeta_2)/z}\ell^{\,2\zeta_2}$ for $\ell\ll\xi$, where $\zeta_2$ is the local roughness exponent, and $G_2=G_2^\mathrm{sat}\propto\tau^{2\beta}$ for $\ell\gg\xi$. When $\zeta=\zeta_2$ the interface has usual or Family-Vicsek scaling and in other cases it has anomalous scaling. 

\section{\!\!Results of the KPZ integration by restricting method\label{sec:3}}

\subsection{Results for 1-dimension\label{subsec:3a}}
We begin by analyzing the results of the simulations with the dimensionless KPZ equation in 1-dimension. We use Euler (or pre-point) discretization and growth restrictions by establishing a maximum height difference around the evolving site.
\begin{figure}[h!]
\centering
\includegraphics[scale=0.3]{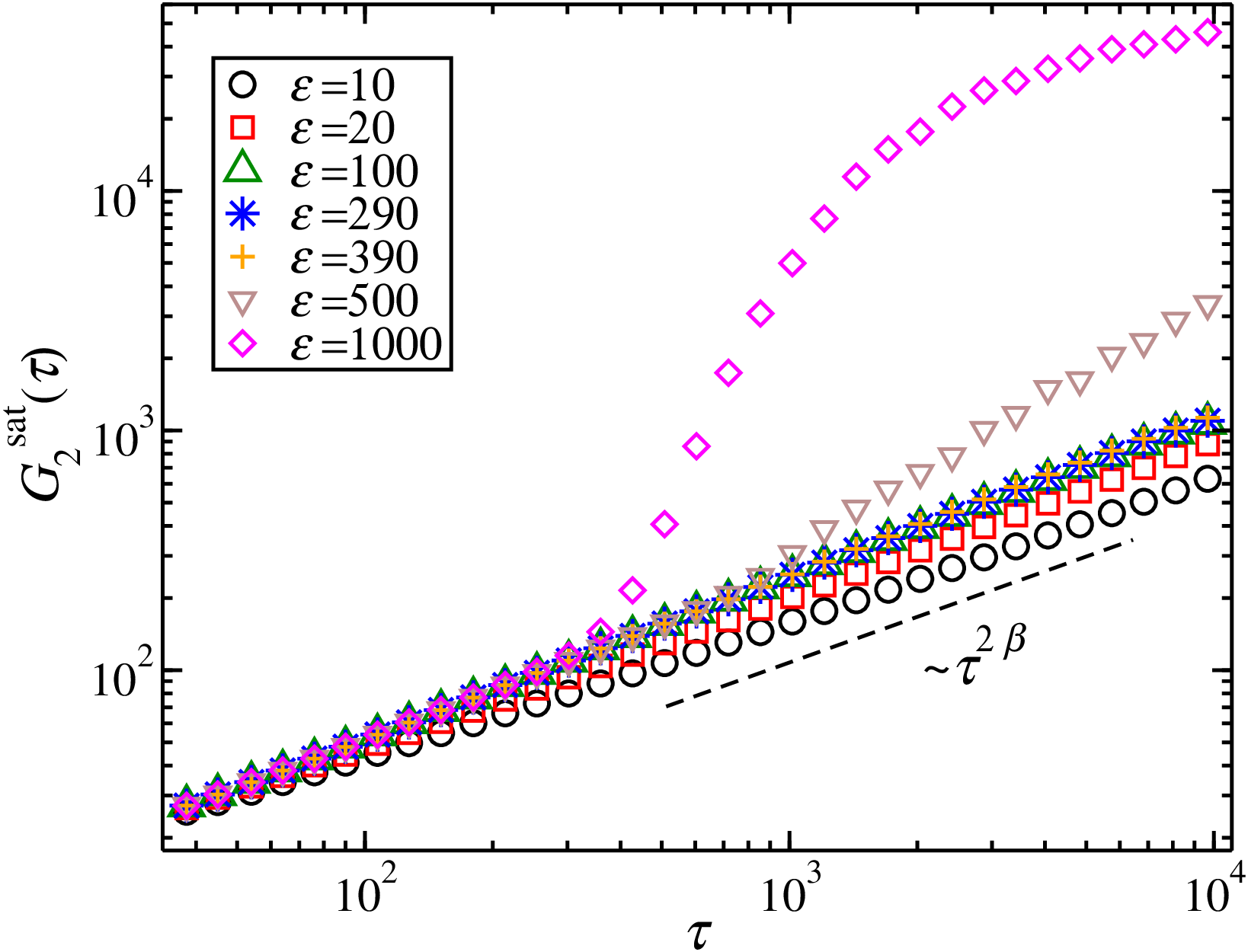}
\caption{(Color online) Height difference correlation of $2^\mathrm{nd}$-order at the saturation $G_2^\mathrm{sat}$ as a function of the adimensional time $\tau$ for several values of $\varepsilon$. We use the same data of Fig.~\ref{Fig:1} and $k=1.5$. The values of the roughness exponent $\beta$ measured (in the cases where it can be measured) are $0.291$ ($\varepsilon=10$), $0.324$ ($\varepsilon=20$), $0.335$ ($\varepsilon=100$), $0.338$ ($\varepsilon=290$) and $0.344$ ($\varepsilon=390$). The plot shows for $\varepsilon=290\approxeq\varepsilon^*(1.5)$ (blue upward-triangle) the power law behavior of the correlation, for four decades, with growth exponent $\beta$ very close to the KPZ theoretical exponent $\beta_\mathrm{KPZ}=1/3$. Below, for $\varepsilon=10$ (black circles) the power law behavior of the roughness exponent is close to the EW theoretical exponent $\beta_\mathrm{EW}=1/4$. Above, for $\varepsilon=500$ (orange left-triangle) the power law behavior, like KPZ, is lost over time due to divergences in integration.\label{Fig:2}}
\end{figure}
Fig.~\ref{Fig:1} shows the plot of probability $\Psi$ as a function of $\varepsilon$ for several values of $k=(\Delta\tau)^{-1}$ and $g=12.56\approxeq 4\pi$. We chose this value because it was originally reported by Moser {\sl et al.} that in a non-dimensionless and non-restricted integration it is the best value for reproducing the KPZ exponents \cite{Moser1991}. The probability $\Psi$ shows two well differentiated behavior as a function of the inverse time step $k$. 

For $k\le k^*$, the function $\Psi$ quickly decreases with $\varepsilon$ to a minimum value at $\varepsilon=\varepsilon^*(k)$ and then increases also rapidly, where $\varepsilon^*$ is a monotonically increasing function of $k$. The minimum of $\Psi$ is visually on the plot for values of \mbox{$k<k^*\approx 2$}. This particular behavior of $\Psi$ can be understood by plotting the height-difference correlation of \mbox{$2^\mathrm{nd}$-order} at the saturation  $G_2^\mathrm{sat}$ as a function of time $\tau$, for different $\epsilon$ values, as shown in Fig.~\ref{Fig:2} for $k=1.5$ and $g=12.56$. Note that $G_2^\mathrm{sat}$ shows the same behavior as the roughness $W$, both as a function of time $\tau$, for the KPZ equation. For $\varepsilon = \varepsilon^*$ a power law behavior of the correlation $G_2^\mathrm{sat}$ with growth exponent very close to the theoretical value of the KPZ growth exponent ($\beta_\mathrm{KPZ} = 1/3$) is observed. For $\varepsilon < \varepsilon^*$ the law of power holds although the exponent moves away as we move away. For $\varepsilon\ll \varepsilon^*$ we can observe that the measured exponent approaches the theoretical value of the EW growth exponent ($\beta_\mathrm{EW} = 1/4$). This change in behavior is a consequence of the nonlinear term being strongly restricted and diffusion being dominant. Conversely, for $\varepsilon > \varepsilon^*$ the behavior of the power law is temporarily shortened with an exponent close to the theoretical value in the early- and middle-time regime and, later, it shows a divergent behavior. This is accentuated when the $\varepsilon$  value is far from $\varepsilon^*$; the $\varepsilon$ value is so large that it is unable to limit divergences of integration. In this last case, the method becomes ineffective and presents the same difficulties as the usual integration method. 

For $k> k^*$, no minimum value is observed for the probability $\Psi(\varepsilon)$. It is simply observed that $\Psi$ is always decreasing as a function of $\varepsilon$, becoming very small or zero when $\varepsilon$ increases, which assures us that the restriction is applied very rarely or never. Even so, the restriction must be maintained, since a divergences in merely one sample would destroy the average. For $k> k^*$ it is enough to stay in the last value of $\varepsilon$ for which $\Psi\neq 0$. Otherwise, for $k=k^*$, we consider the value $\varepsilon=\varepsilon^*$ where $\Psi$ has a minimum.

\begin{figure}[h!]
\centering
\includegraphics[scale=0.3]{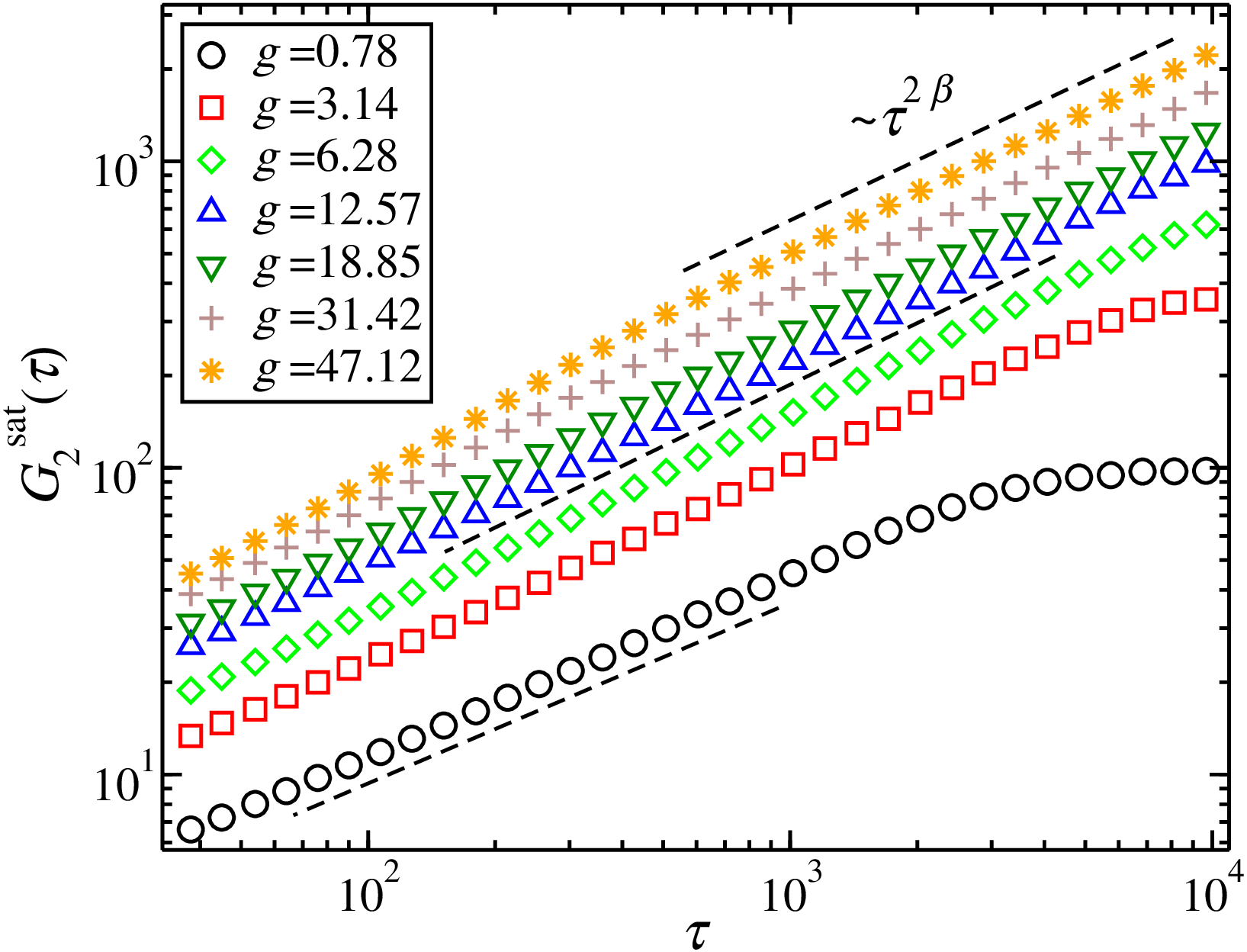}
\caption{(Color online) Saturation of the $2^\mathrm{nd}$-order height-difference correlation $G_2^\mathrm{sat}$ as a function of time $\tau$ for several values of $g$. For the simulations we used $L = 512$, $k = 100$ and $\varepsilon$ chosen as follows: $\varepsilon = 200$ for $g<4\pi$ and $g = 300$ otherwise.
The values of the roughness exponent $\beta$ measured  are $0.298$ ($g=0.78\approx\pi/4$), $0.324$ ($g=3.14\approx\pi$), $0.330$ ($g=6.28\approx 2\pi$), $0.331$ ($g=12.56\approx 4\pi$), $0.333$ ($g=18.85\approx 6\pi$), $0.330$ ($g=31.42\approx 10\pi$) and $0.329$ ($g=47.12\approx 15\pi$).\label{Fig:3}}
\end{figure}

To observe how the growth exponent $\beta$ depends on the coupling parameter $g$ we study the correlation $G_2^\mathrm{sat}$ as a function of time $\tau$ for different values of $g$. Taking $L = 512$ and $k = 100$, in Fig.~\ref{Fig:3} we observe how the power laws are maximized in the neighborhood of $g = 15$. For values of $g\ll 15$ the elastic term causes a rapid saturation of the interface, producing a small deviation from the expected KPZ-value of $\beta$. For $g\gg 15$ the noise term affects the growth at the beginning, delaying the KPZ behavior. Both behaviors are due to finite-size effects, {\sl i.e.} for all $g$, as the system size grows, $\beta\to\beta_\mathrm{KPZ}$.

Results of our simulations show that the system is self-affine. The \mbox{$m^\mathrm{th}$-order} correlation $G_m$ as a function of the distance $\ell$ between columns show that $\zeta_1\approxeq 0.487$, $\zeta_2\approxeq 0.489$, $\zeta_3\approxeq 0.488$, and $\zeta_4\approxeq 0.489$ for a system of size $L=16384$, $k=10$ and and time $\tau=2^{20}$ close to the saturation.  The exponent values in different orders are very close to each other, which confirms that the restrictions do not introduce changes to the results known for the KPZ. Simulations show that the system with restriction maintains the self-affinity for different sizes $L$ and for different values of the coupling constant $g$. 

\begin{figure}[h!]
\centering
\includegraphics[scale=0.3]{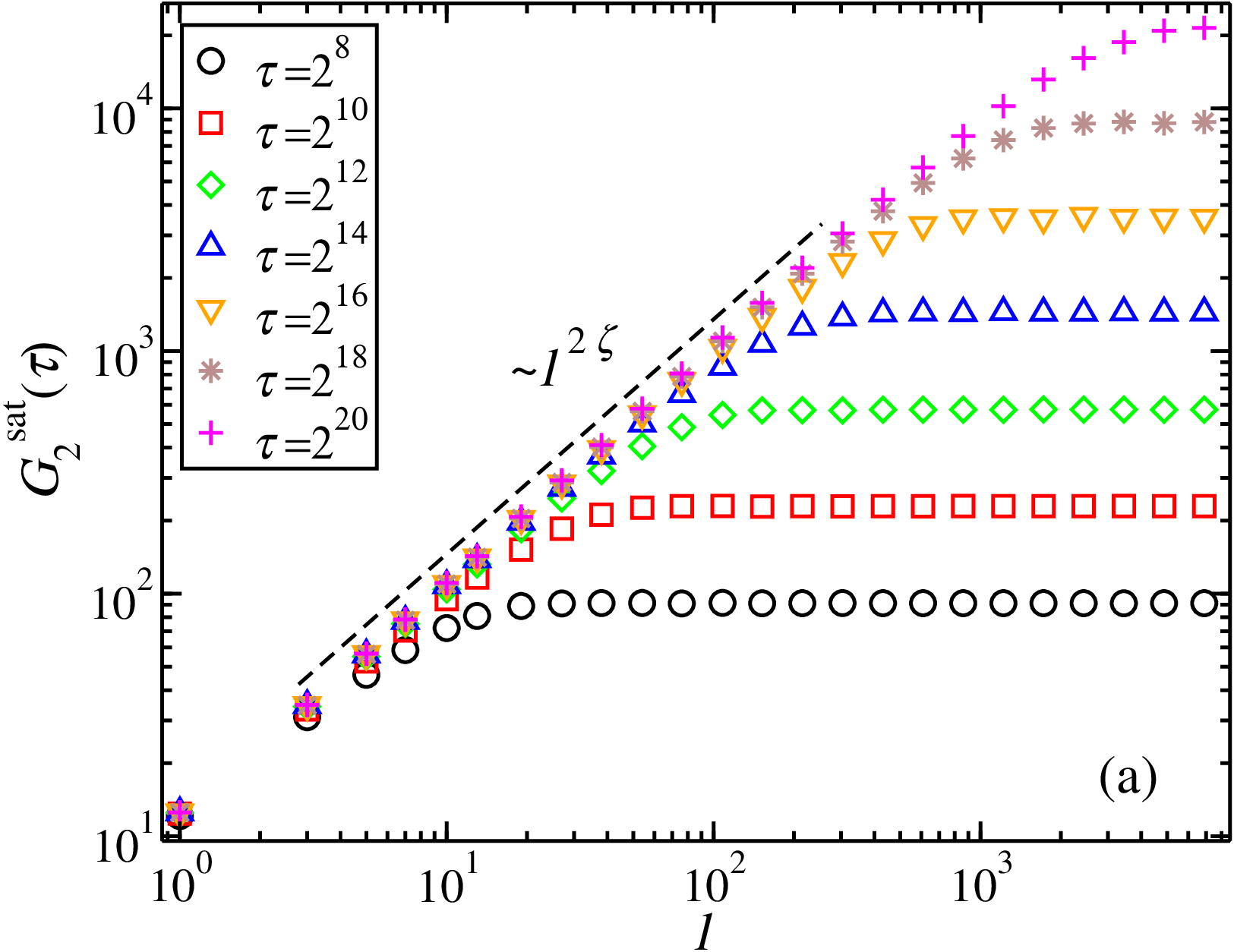}\\
\includegraphics[scale=0.3]{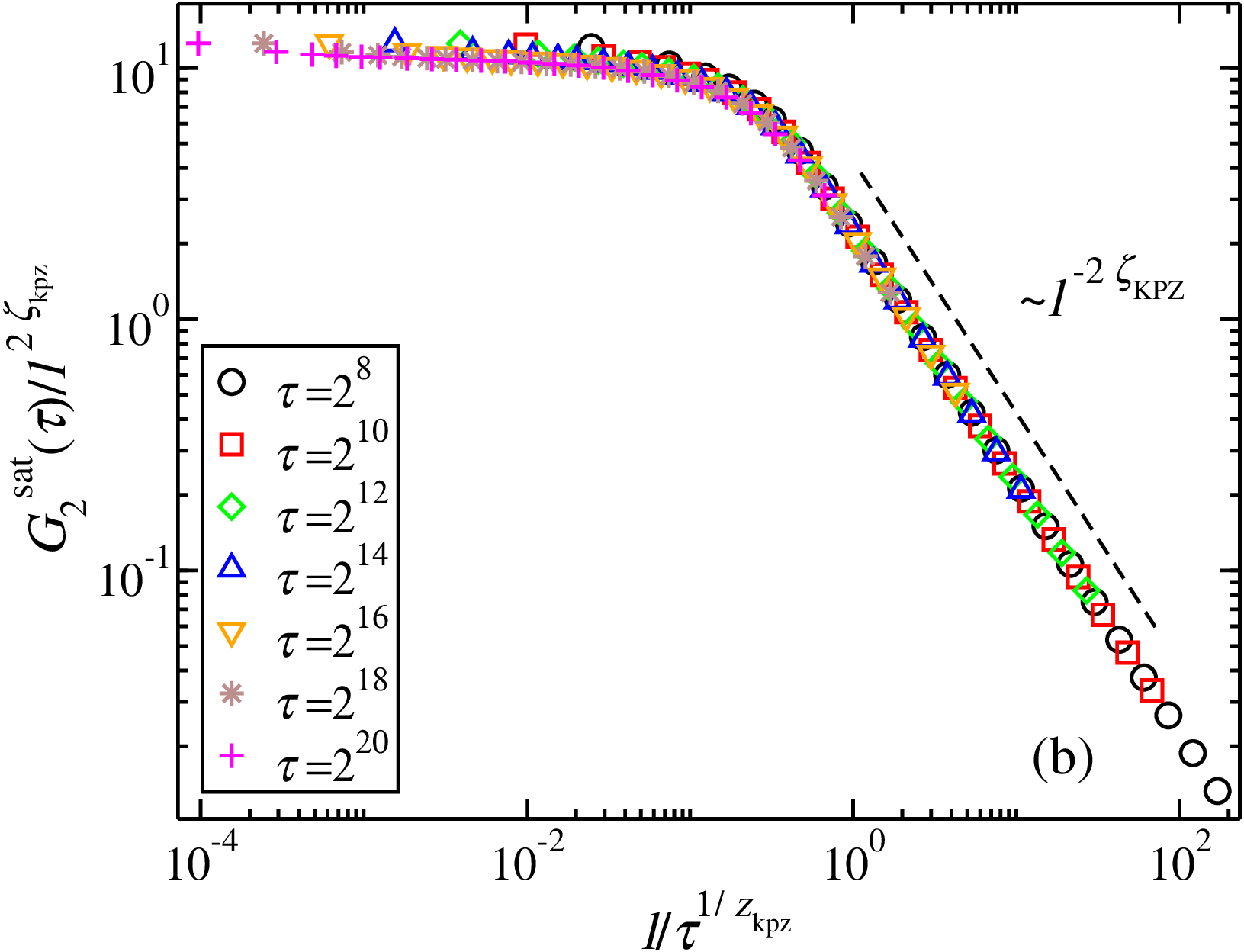}
\caption{(Color online) (a) Height-difference correlation of $2^\mathrm{nd}$-order $G_2$ as a function of the distance between columns $\ell$ for several times $\tau=2^n$ (with $n$ integer), taking $L=16384$, $k=100$, $g=12.56$, and $\varepsilon=250$. The values of the local roughness exponent $\zeta_2$ measured are $0.440$ ($n=10$), $0.457$ ($n=12$), $0.471$ ($n=14$), $0.479$ ($n=16$), $0.485$ ($n=18$) and $0.489$ \mbox{($n=20$)}. As $\tau$ increases $\zeta_2$ approaches the KPZ theoretical value of the global roughness exponent \mbox{$\zeta_\mathrm{KPZ}=1/2$}. (b) Scaling of the $G_2(\ell,\tau)$ for the same parameters and values of $\tau$ as those used in the plot (a). The plot shows $G_2=\const$ for $\ell\ll\tau^{1/z}$ and $G_2\propto\tau^{2\beta}$ for $\ell\gg\tau^{1/z}$.\label{Fig:4}}
\end{figure}

Plot (a) of Fig.~\ref{Fig:4} shows $G_2$ as a function of $\ell$, for different values of time $\tau$ and size $L=16384$ and $k=10$. We observe that $\zeta_2$ depends on $\tau$ and  it approaches the KPZ theoretical value ($\zeta_{\mathrm{KPZ}}=1/2$) as the time $\tau$ increases. As other authors have shown \cite{Miranda2008}, when the system size $L$ and time $\tau$ got $+\infty$ the global roughness exponent $\zeta_2(\tau)\to 1/2$. Because the correlation lenght $\xi\propto \tau^{1/z}$  we plot in Fig.~\ref{Fig:4} (b)  the scaling function $\ell^{-2\zeta_\mathrm{KPZ}} G_2$ as a function of $\ell\,\tau^{-1/z_\mathrm{KPZ}}$ for several values of $\tau$. From the plot we can see that the interface has usual scaling, since the different curves overlap. It is important to note that, at distance $\ell_0\approx 2$, the correlation $G_2(\ell_0,\tau)$ is constant with time $\tau$, except from small initial variations (see Fig.~\ref{Fig:4}). Therefore, if we increase the size of the system, bringing it to the thermodynamic limit, it is not necessary to modify the restriction that allows the integration. In contrast, as the system size increases, our method for growth models with anomalous scaling ({\sl e.g.} LD equation) requires modifying the restriction that avoids instabilities. Similarly, the method of integration of Dasgupta {\sl et al.} for the LD equation shows that by increasing the size of the system it is necessary to modify the parameter $c$ that controls the nonlinearities that avoid instabilities \cite{Dasgupta1996, Dasgupta1997}.

Similar results are observed when the KPZ Eq.~(\ref{eq:KPZ}) includes an additive term representing a constant force $f\neq 0$ due to incoming or outcoming particle-flow. As $f$ increases the only relevant change is an initial deformation in the roughness similar to incresing $g$ in Fig.~\ref{Fig:3}. 

\subsection{Results for $d$-dimensions ($d=3,4$)\label{subsec:3b}}

\begin{figure}[h!]
\centering
\includegraphics[scale=0.3]{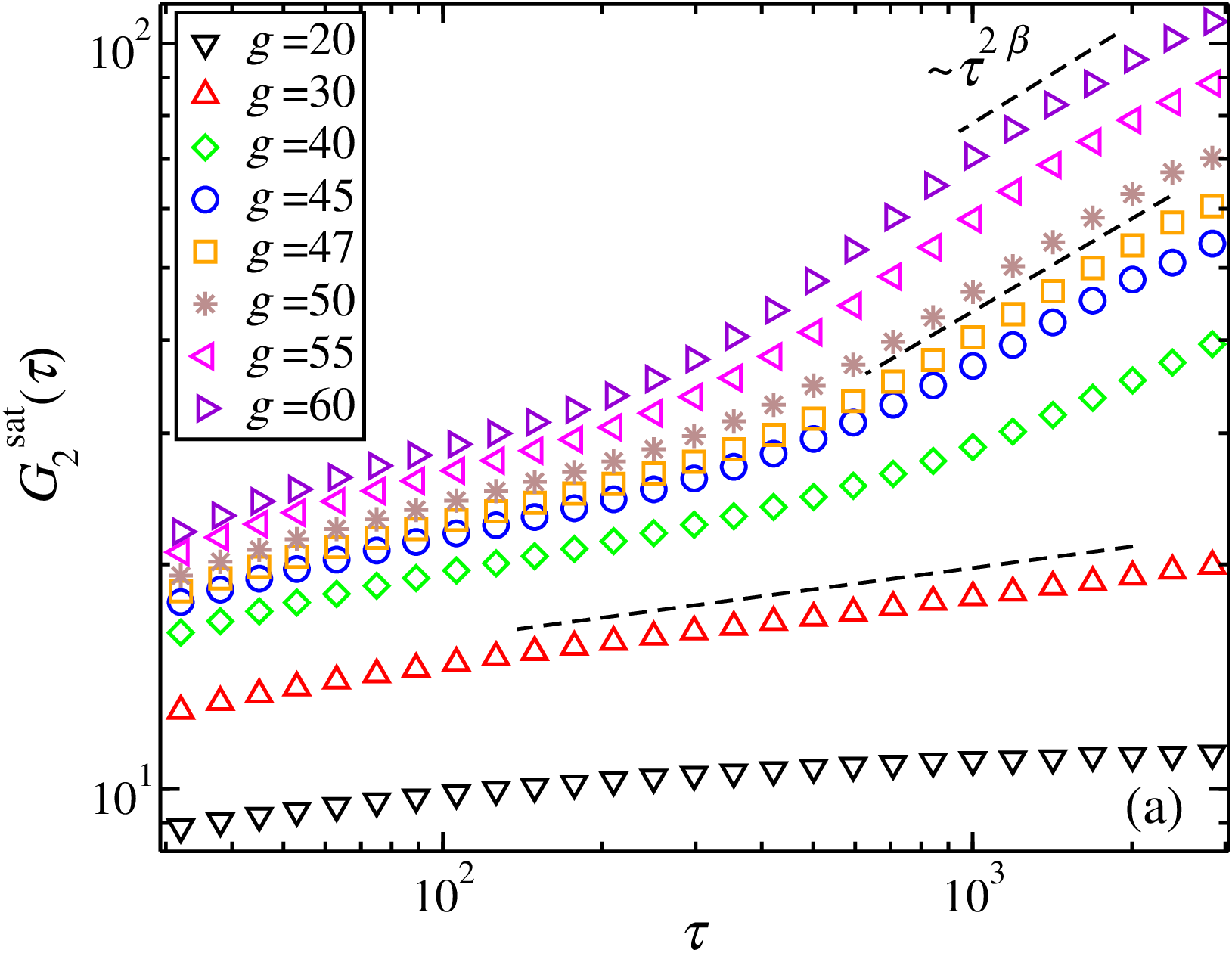}\\
\includegraphics[scale=0.3]{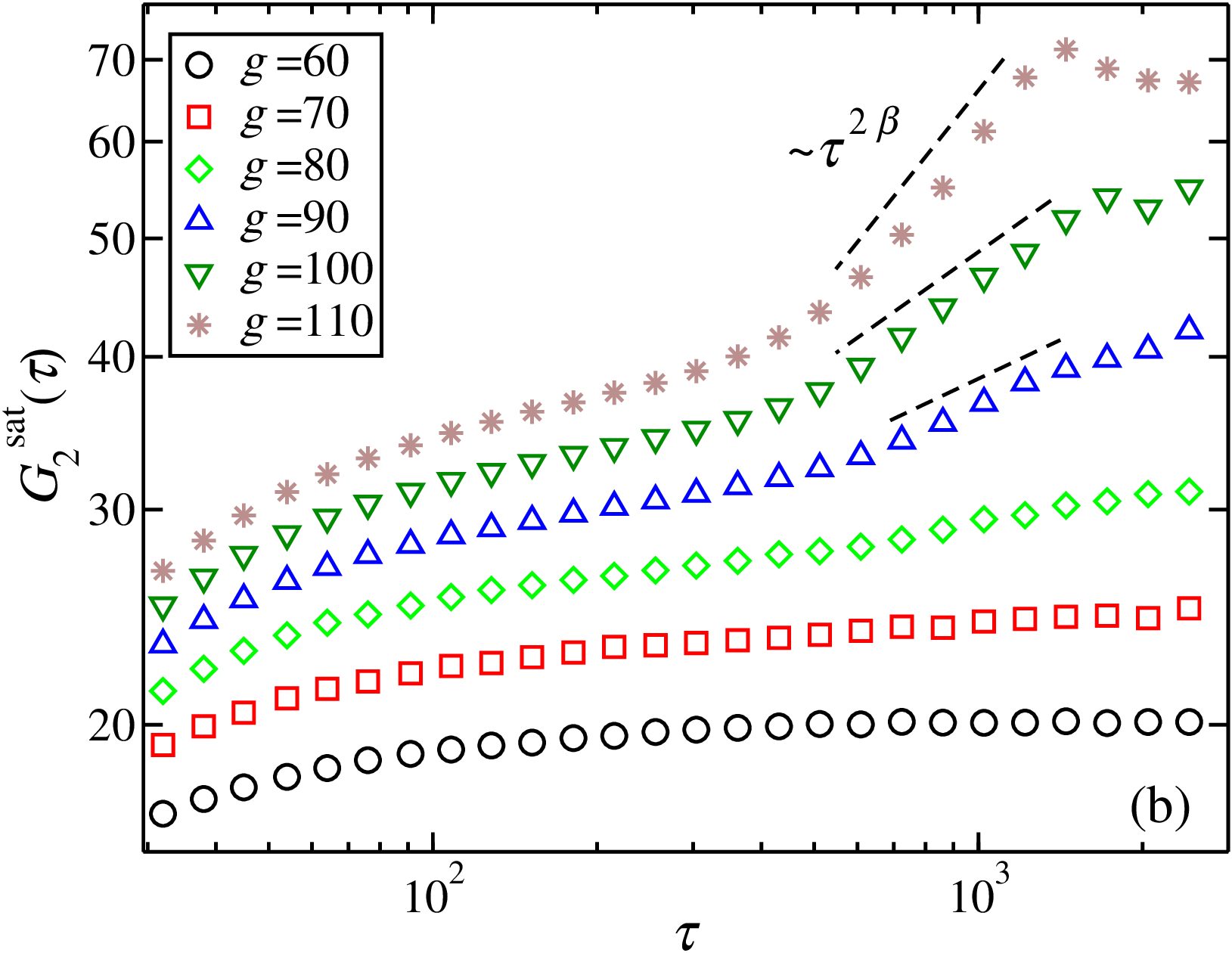}
\caption{(Color online) Both plots: Height-difference correlation of $2^\mathrm{nd}$-order $G_2^\mathrm{sat}$ at the saturation as a function of the time $\tau$ for several $g$. For $d=3$ (Plot (a)) we take $L=60$, $k=50$ and $\varepsilon=130$.  The values of the roughness exponent $\beta$ measured (in the cases where it can be measured) are $0.045$ ($g=30$),  $0.156$ ($g=40$),  $0.193$ ($g=45$),  $0.204$ ($g=47$),  $0.220$ ($g=50$),  $0.198$ ($g=55$) and  $0.201$ ($g=60$). For $d=4$ (Plot (b)) we take $L=16$, $k=100$ and $\varepsilon=130$. The values of the roughness exponent $\beta$ measured (in the cases where it can be measured) are $0.106$ ($g=90$),  $0.159$ ($g=100$) and  $0.273$ ($g=110$).\label{Fig:5}}
\end{figure}
By applying the methodology tested in the 1-dimensional case we will analyse some important results that can be reached in $d$-dimensions ($d>2$). In particular, we will consider the cases $d=3,4$. For $d=3$ the height difference correlation of $2^\mathrm{nd}$~order at the saturation $G_2^\mathrm{\!sat}$  shows that there is a critical coupling value $g_\mathrm{c}$ that separates two well-differentiated behaviors. Below (\mbox{$g<g_\mathrm{c}$}) the weak-coupling phase is observed, where $\beta\approx 0$ and the system saturates rapidly. For $g\approx g_\mathrm{c}$, since it is a critical point, a power law-like behavior is observed, with $\beta\gtrapprox 0$. Above (\mbox{$g>g_\mathrm{c}$}), when $g$ increases the interface roughness begins to increase. For a fixed value $g=g_\mathrm{o}$, the optimal power-law behavior of the strong-coupling phase is obtained. For $g>g_\mathrm{o}$, it is observed, as for $d = 1$, that the noise initially affects the power law, delaying the appearance of the KPZ type behavior, but maintaining the exponents approximately. In Fig.~\ref{Fig:5} (a) the correlation $G_2^\mathrm{ sat}$ is plotted as a time-dependent function $\tau$ for $d=3$ at different values of $g$. We observe that $g_\mathrm{c}\approx 30$ measuring $\beta(30)\approxeq 0.045$ and that the value of $g$ for which the best power law is obtained is close to $g_\mathrm{o}\approx 47$, with $\beta(47)\approxeq 0.204$. For higher values ​​of $g$ the exponent value is $\beta\gtrapprox 0.203$, except for $g = 50$, where a greater deviation occurs, reaching $\beta(50)\approxeq 0.22$. We believe that this deviation is due to the fact that the noise modifies the power law smoothly, before transforming into the deformation seen for larger $g$ values. However, an exponent value $0.203\lessapprox\beta\lessapprox 0.220$ is very close to that measured for the RSOS model \cite{AlaNissila1993}. Taking $\beta= 0.203$ and accepting the relation $\zeta + z = 2$ (with $z=\zeta/\beta$) we obtain $z\approxeq 1.662$ and $\zeta\approxeq 0.337$. For the RSOS model as well, the roughness exponent $\zeta\approxeq 0.313$ was measured \cite{Marinari2002}, a value with a small departure from those mentioned above.

For the case of $d=4$ the correlation $G_2^\mathrm{sat}$ is plotted as a function of time $\tau$ for different values of $g$ on the (b) of Fig.~\ref{Fig:5}. A behavior similar to the case $d=3$ is observed, with $g_\mathrm{c}\approx 80$ and $g_\mathrm{o}\approx 100$, measuring $\beta(100)\approxeq 0.159$. This value is very close to that measured for RSOS in the same dimension \cite{Kim2013}. Taking the recently measured exponent $\zeta\approxeq 0.273$ for the RSOS model \cite{Pagnani2013, Kim2013}, applying the relation $\zeta/\beta+\zeta=2$, we calculate $\beta\approxeq 0.158$, a value very close to that reported here. For $g>100$, the measured values for the $\beta$ exponent begin to deviate significantly from the values accepted and retrieved here. We observe that because the system size is very small, for $d = 4$ it saturates very fast and we cannot conclude if the KPZ behavior is recovered as for $d = 1$ and $d = 3$.

\section*{Conclusions}

In this paper, we demonstrate how it is possible to avoid the instabilities associated with the uncontrollable growth of pillars or wells that can be developed during the evolution of the discrete version of the KPZ equation. The method we propose here does not modify the discrete KPZ equation by adding nonlinear terms, as the proposal introduced 20 years ago by Dasgupta {\sl et al.}. In contrast, we propose to impose restrictions to the lateral growth of the interface by limiting the nonlinearities of the KPZ equation. The restriction rule is applied at each integration time step, but only acts to eliminate divergences while maintaining all the properties of the continuous KPZ equation. The $\varepsilon$ restriction parameter is chosen in a range of values that leave the scaling properties invariant. In this work, we have integrated the discrete version of the dimensionless KPZ equation in such a way that it only depends on two parameters, aside from $\varepsilon$: the inverse of the time-step $k$ and the coupling constant $g$. Our results in 1-dimension, with Euler discretization and growth restrictions, show that the height-difference correlation of $2^\mathrm{nd}$-order at the saturation is a power law with a growth exponent very close to the theoretical value established for the KPZ equation, {\sl i.e.} $\beta_\text{\tiny KPZ}=1/3$. The method becomes effective for all times under these condition. Otherwise, if $\varepsilon\ll\varepsilon^*$ the growth exponent is close to the theoretical value of the EW universality class ($\beta_\text{\tiny EW}=1/4$) and if $\varepsilon\gg\varepsilon^*$ the KPZ power law can be broken due the emergence of divergences with a probability that decreases with $k$. The method was tested for $d=3$ and $d=4$. In the first case, the predicted result is obtained by numerical methods, showing both the weak- and strong-coupling phases. Our results yield a critical-coupling constant $g_\mathrm{c}\approx 30$ and an optimal-coupling constant $g_\mathrm{o}\approx 47$ of the strong-coupling phase, for which the exponent $\beta$ is close to that measured for RSOS simulations. When integrating at $d=4$, a strong-coupling phase is observed for $g>g_\mathrm{c}\approx 80$ with an exponent close to that measured for RSOS at the optimal-coupling constant $g_\mathrm{o}\approx 100$. This can be taken as an indication that $d=4$ is not the critical dimension of the KPZ universality class. Another option is that the observed strong-coupling phase is attributed to the finite size of the sytem. However, in our simulations, increase the size of the system does not result in the disappearance of this phase.

\begin{acknowledgements}
This work was partially supported by Consejo Nacional de Investigaciones Cient{\'i}ficas y T{\'e}cnicas (CONICET), Argentina, PIP 2014/16 No. 112-201301-00629. R.C.B. thanks C. Rabini for her suggestions on the final manuscript.
\end{acknowledgements}

\bibliography{TB-KPZ-intg.bib}

\end{document}